\documentclass[aps,prl,twocolumn]{revtex4}

\usepackage{amssymb,amsmath,mathrsfs,epstopdf,slashed,color}

\usepackage{ifpdf}
\ifpdf
\usepackage{graphicx}
\usepackage{hyperref}
\else
\usepackage[dvipdfmx]{graphicx}
\usepackage[dvipdfmx]{hyperref}
\usepackage[numbers]{natbib}
\usepackage{hypernat}
\fi

\DeclareMathOperator*{\re}{Re \,}
\DeclareMathOperator*{\im}{Im \,}

\begin{document}


  



  
 




  
 
 


\title{Preference for a Vanishingly Small Cosmological Constant \\
in Supersymmetric Vacua in a Type IIB String Theory Model
}
\author{Yoske Sumitomo${}^{1}$}
\author{S.-H. Henry Tye${}^{1,2}$}

\affiliation{
 ${}^1$ Institute for Advanced Study, Hong Kong University of Science and Technology, Hong Kong\\
 ${}^2$ Laboratory for Elementary-Particle Physics, Cornell University, Ithaca, NY 14853, USA
}

\date{\today}

\begin{abstract}
  We study the probability distribution $P(\Lambda)$ of the cosmological constant $\Lambda$ in a specific set of KKLT type models of supersymmetric IIB vacua. We show that, as we sweep through the quantized flux values in this flux compactification, 
 $P(\Lambda)$ behaves divergent at $\Lambda =0^-$ and the median magnitude of $\Lambda$ drops exponentially as the number of complex structure moduli $h^{2,1}$ increases.
 Also, owing to the hierarchical and approximate no-scale structure, the probability of having a positive Hessian (mass squared matrix) approaches unity as $h^{2,1}$ increases.
\end{abstract}

\maketitle


\section{Introduction}

The cosmological constant $\Lambda$ as the dark energy is exponentially small; setting Planck scale $M_P=1$, we have 
\cite{Weinberg:1987dv,Bennett:2012fp} (and references therein),
\begin{equation}
 \Lambda \sim 10^{-122}.
\end{equation}
Since string theory has many solutions (i.e., leading to the so called cosmic landscape), we are interested in the probability distribution $P(\Lambda)$ of the cosmological constant $\Lambda$ of the meta-stable solutions. In this paper, we consider a class of supersymmetric vacua (of the well studied KKLT type) in Type IIB string theory \cite{Kachru:2003aw,Giddings:2001yu}.

At low energies, we obtain an effective supergravity theory after 6 of the 9 spatial dimensions are compactified. 
A typical flux compactification in string theory involves many moduli and $3$-form field strengths with quantized fluxes. Together with the quantized fluxes, the moduli and their dynamics describe the string theory landscape. It is pointed out \cite{Bousso:2000xa} that the spacing $\delta \Lambda$ can be exponentially small, so a very small $\Lambda$ value is possible. However, this alone does not explain why nature picks such a very small $\Lambda$, instead of a value closer to the string or Planck scale.

Consider the 4-dimensional low energy supergravity effective potential $V$ for the vacua coming from some flux compactification in string theory. To be specific, let us consider only $3$-form field strengths $F^I_3$ wrapping the 3-cycles in a Calabi-Yau type manifold. So we have
\begin{equation}
 \begin{split}
  & V( F^I_{3}, \phi_j) \rightarrow V(n_i, \phi_j), \\
  & i=1,2,...,N, \quad {\rm and} \quad  j=1,2,..., K
 \end{split}
\end{equation}
where the flux quantization property of the $3$-form field strengths $F_3^I$ allow us to rewrite $V$ as a function of  the quantized values $n_i$ of the fluxes present and $\phi_j$ are the moduli describing the size and shape of the compactified manifold as well as the coupling. There are finite barriers between different  sets of flux values.

For a given set of ${n_i}$, we can solve $V(n_i, \phi_j)$ for its meta-stable (classically stable) vacuum solutions via finding the values $\phi_{j, {\rm min}} (n_i)$ at each solution and determine $\Lambda=\Lambda (n_i)$.  Collecting all such solutions, we can next find the probability distribution $P(\Lambda)$ of $\Lambda$ of these meta-stable solutions as we sweep through the flux numbers $n_i$. A typical $n_i$ can take a large range of integer values ($|n_i| \le M_P^2/M_s^2$), subject to validity constraints. 
So we may simply treat each $n_i$ as a random variable with some uniform distribution $P_i(n_i)$ and, given $\Lambda (n_i)$, we can determine $P(\Lambda)$. With $P(\Lambda)$, we can find both the average and the median value of $\Lambda$. Since the ranges of $n_i$ can be very large, we may approximate them as continuous random variables.

It turns out  that the string theory dynamics
(i.e., the resulting functional form of $\Lambda (n_i)$)
together with simple probability theory typically yields a $P(\Lambda)$ that peaks (i.e., diverges) at  $\Lambda=0$. In fact, this peaking at $\Lambda=0$ behavior is relatively insensitive to the details of the
smooth distributions $P_i(n_i)$ and  becomes more divergent as the number of moduli and fluxes increase. This divergence is always mild enough so $P(\Lambda)$ can be properly normalized,
i.e., $\int P(\Lambda) d \Lambda=1$, henceforth implying that the probability at exactly $\Lambda=0$ will remain exactly zero. Since the number of moduli as well as cycles that fluxes can wrap over in a typical known flux compactification (e.g., the well studied ${\mathbb CP}^4_{11169}$ has 272 complex structure moduli \cite{Candelas:1989js,Denef:2004dm}) is of order ${\cal O}(100)$, a vanishingly small but non-zero $\Lambda$ appears to be statistically preferred.

To be specific, the model of interest in this paper is given by its 4-dimensional low energy effective supergravity approximation,
\begin{align}
 V =& e^K \left(|D_T W|^2 + |D_S W|^2 + |D_{U_i} W|^2 - 3 |W|^2 \right),\notag \\
 K =& -3 \ln \left(T+\bar{T}\right) - \ln \left(S+\bar{S}\right) - \sum_{i=1}^{h^{2,1}} \ln \left(U_i + \bar{U}_i \right), \notag \\
 W =& W_0 + A e^{-a T},\label{total potential} \\
   W_0 =& c_1 + \sum_{i=1}^{h^{2,1}} b_i U_i - \left( c_2 + \sum_{i=1}^{h^{2,1}} d_i U_i \right) S, \notag 
\end{align}
where $T$ is the K\"ahler modulus (the field that measures the volume of the 6-dimensional manifold), $S$ is the dilation and $U_i$ ($i=1, 2, ..., h^{2,1}$) are the complex structure (or shape) moduli.  
Note that $W_0$ in general takes the form \cite{Gukov:1999ya},
\begin{equation}
\label{Ccoef}
W_0(U_i,S) =  \sum_{cycles}  \int G_3 \wedge \Omega =C_1(U_i) + S C_2(U_i),
\end{equation}
where $G_3 =F_3 - iSH_3$ and $F_3=dC_2$ is the Ramond-Ramond (RR) type  and $H_3=dB_2$ is the Neveu-Schwarz (NS-NS) type. 
Here $C_i(U_i)$ are complex functions of $U_i$ only. Expanding them in terms of the flux parameters,
\begin{equation}
C_1(U_i)=  c_1  + b_iU_i +  \cdots ,  \quad C_2(U_i)=  c_2 + d_iU_i + \cdots . 
\end{equation}
For general compactifications, $W_0 (U_i, S)$ remains to be determined. 
In orientifolded toroidal orbifolds, the complex flux parameters take the form 
\begin{equation}
b_i  \propto n_{bi} + \omega_{bi} m_{bi},  \quad n_{bi}, m_{bi} \in {\bf Z}
\end{equation}
where $|\omega_{bi}|=1$
and similarly for $c_i$ and $d_i$
 \cite{Lust:2005dy,Rummel:2011cd}. So the discrete flux parameters $c_i$, $b_i$ and $d_i$ may be treated as independent random variables with uniform distributions. 

Since the quantized fluxes of the 3-form field-strengths are expected to vary over large ranges \cite{Bousso:2000xa,Denef:2004ze,Denef:2004cf}, the discrete flux parameters $b_i$, $c_j$ and $d_i$ may be treated as continuous parameters sweeping through some smooth uniform ranges. Validity of the weak coupling approximation requires that $s= \re S >1$ while reality of the K\"ahler potential $K$ requires that $u_i=\re U_i >0$. The second term (with parameters $a>0$ and $A$) in the superpotential $W$ is a non-perturbative term emerging from, e.g., gaugino condensation.

\begin{figure}
 \begin{center}
  \includegraphics[width=24em]{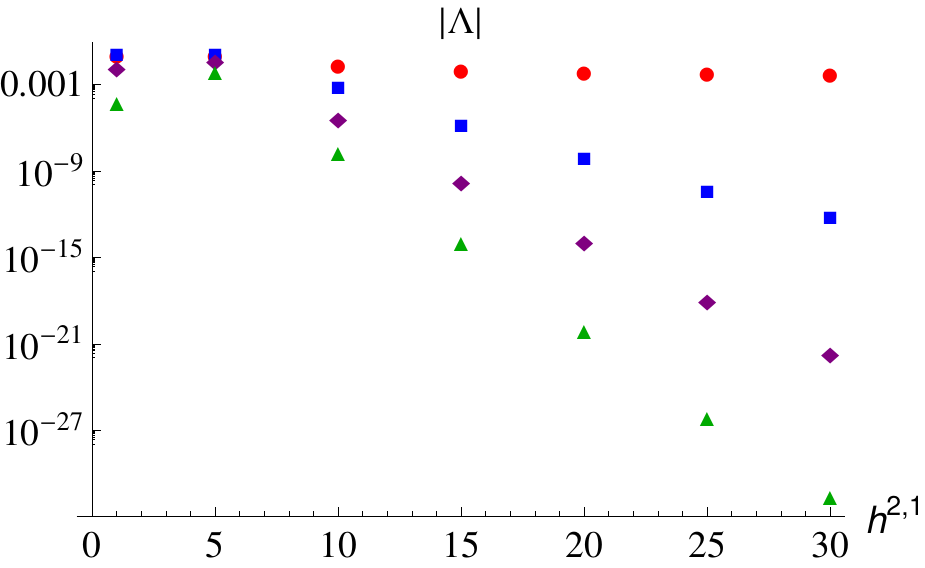}
 \end{center}
 \caption{The comparison of $\left< |\Lambda| \right>$ (red circle), $|\Lambda|_{80\%}$ (blue square), $|\Lambda|_{50\%}$ (purple diamond), $|\Lambda|_{10\%}$ (green triangle) are shown as functions of the number of complex structure moduli $h^{2,1} = 1, 5, 10, 15, 20, 25, 30$. }
 \label{fig:k+cs-distribution}
\end{figure}

After finding a supersymmetric vacuum solution, we can express $\Lambda$ in terms of the physical parameters. Since there are many solutions, the values of $\Lambda$ are expected to form a closely spaced "discretuum". 
Treating the flux parameters as random variables with (smooth) distributions $P_i(b_i), P_i(d_i), P_i(c_i)$, we are able to find the probability distribution $P(\Lambda)$.
\begin{itemize} 

\item $P(\Lambda)$ peaks at $\Lambda=0^-$ for the model (\ref{total potential}) with multi-complex structure moduli. As an illustration, we simplify the problem by treating the flux parameters to be real random variables. figure \ref{fig:k+cs-distribution} shows the result of a specific set of uniform distributions for the flux parameters. Here, the expected value $\left< |\Lambda| \right>$ of $\Lambda$ is shown as a function of the number $h^{2,1}$ of complex structure moduli. Because of the long tail in $P(\Lambda)$, we see that  $\left< |\Lambda| \right>$ does not drop as $h^{2,1}$ increases. To get a better feeling of the property of $P(\Lambda)$, let us introduce $|\Lambda|_{Y\%}$, defined by
$\int^0_{-|\Lambda|_{Y\%}} \, P(\Lambda) \, d\Lambda = Y\%$. That is, there is  a $Y\%$ probability that $\Lambda$ will fall in the range $0 \le |\Lambda| \le |\Lambda|_{Y\%}$. 
In figure \ref{fig:k+cs-distribution}, we see that a typical $|\Lambda|_{Y\%}$ decreases exponentially as $h^{2,1}$ increases. If we take the median value $|\Lambda|_{50\%}$ to be the likely value of $|\Lambda|$ and assume its behavior in figure \ref{fig:k+cs-distribution} extends to larger $h^{2,1}$, we see that, for $h^{2,1}>5$,
\begin{equation}
|\Lambda| \simeq 10^{-0.82 \, h^{2,1} + 2.7}.
\end{equation}
Since $h^{2,1} \sim {\cal O}(100)$ in some known models (see e.g. \cite{Denef:2004dm} and references therein), we see that it is quite natural to get a vanishingly small $\Lambda$.

\item To prepare for the analysis of the full model (\ref{total potential}), we study analytically the single K\"ahler modulus case. In this simple case, $P(\Lambda)$ (shown in figure \ref{fig:comparison-lambda}) already diverges at $\Lambda=0^-$: $P(\Lambda) \simeq (\ln |\Lambda|)^2 / 2 e^{3/2} \sqrt{|\Lambda|}$, where only the parameter $W_0=c_1$ is treated as a random variable with a uniform distribution.
      It would be interesting to compare with the peaking behavior of $P(\Lambda)$ in the K\"ahler Uplift models where two random parameters are required to have a log divergence as a result of the product distribution \cite{Sumitomo:2012wa,Sumitomo:2012vx}.

\item Another interesting property is the positivity of the Hessian ($\partial_i \partial_j V$). After making the standard assumption that the axionic modes will receive periodic potentials that stabilize them, we may focus on the real components of the moduli only. Here we see that the probability of having a positive Hessian increases quickly towards unity as $h^{2,1}$ increases. This property is due to the approximate no-scale behavior and the hierarchical structure adopted. The latter is fully justified for large $h^{2,1}$.

\end{itemize}

This statistical preference for a vanishingly small $\Lambda$ 
is interesting since energetics alone may suggest that more negative $\Lambda$ should be preferred. Here we see that this  small $|\Lambda|$ property is actually a consequence of the probability theory for the specific functional form of $\Lambda$ in terms of the flux parameters  \cite{Sumitomo:2012wa}. This result strengthens our earlier observation for semi-positive $\Lambda$, where, because of supersymmetry breaking, the functional form of $\Lambda$ is only known approximately \cite{Sumitomo:2012vx}.  Combining these results, we see that there are string theory scenarios where a vanishingly small $\Lambda$, either positive or negative, is generic.  
That the Hessian is mostly positive allows us to be optimistic in the search for de-Sitter vacua in this type of models.

\section{The Single K\"ahler Modulus Case}

To prepare for a full analysis, we first focus on the simple model for a single K\"ahler modulus stabilization at supersymmetric vacua as in \cite{Kachru:2003aw},
\begin{equation}
 \begin{split}
  K_{K}=-3 \ln \left(T+\bar{T}\right), \quad
  W=c_1 + A e^{-a T}.
  \label{K1M}
 \end{split}
\end{equation}
where $W_0=c_1$ here.
Now we solve the supersymmetric {condition $D_T W = \partial_T W + (\partial_T K) W=0$.
The imaginary part of $T$ would have a cosine type of potential and therefore the solution stays at $\im T = 0$.
Then the condition becomes
\begin{equation}
 \begin{split}
  -3y=(2x+3) e^{-x}
 \end{split}
 \label{susy condition}
\end{equation}
where we defined $x=at= a \re T$ and the parameter $y=c_1/A$.
The real part of the modulus is defined to be positive so that the volume of compactification is correctly measured.
Therefore the RHS of (\ref{susy condition}) is not only positive, but has an upper bound value of $3$.
The corresponding condition for the LHS implies $-1 < y \leq 0$.
If we rewrite in terms of $X=-x - 3/2$, (\ref{susy condition}) becomes
\begin{equation}
z \equiv  {3\over 2} e^{-3/2} y = X e^X.
\end{equation}
where $z \simeq 0.3y$. The solution of this equation is known as a {\it Lambert} ${\cal W}$ function or a product logarithm.
There are two branches of solutions : ${\cal W}_{-1}$ for $X \le -1$ and ${\cal W}_{0}$ for $X \ge -1$.
Since $t$ measures the volume of the 6-manifold for compactification, we need $X<-3/2$.
This branch of solution can be expanded around $z\lesssim 0$ by
\begin{equation}
 \begin{split}
 X(y)={\cal W}_{-1} \left(z\right)
  = \ln (-z) -\ln[-\ln (-z)] +  \cdots.
 \end{split}
 \label{solution for T}
\end{equation}
So $x=-(X+3/2)$ is solved in terms of $y$.
Inserting this into the potential $V = e^{K_K} (K_K^{T\bar{T}}|D_T W|^2 - 3 |W|^2)$ sitting at a minimum, we get
\begin{equation}
 \begin{split}
  \Lambda &\equiv V|_{\rm min} = -3 e^{K_K} |W|^2 \\
  =& a^3 A^2 e^{3+2 {\cal W}_{-1} (z)} /  \left[9+6 {\cal W}_{-1} \left(z \right) \right] \\
  &\stackrel{y \rightarrow 0} {\simeq} a^3 A^2 {3 y^2 \over 8 \ln (-y)}.
 \end{split}
 \label{Lambda}
\end{equation}
We see that small $\Lambda$ is precisely related to small $y$ or $c_1$.

Using the condition (\ref{susy condition}) to eliminate $A$, we obtain
\begin{equation}
 \begin{split}
  \partial_{t}^2 V|_{\rm min} = {3 a^5 W_0^2} {2 x^2 + 5 x + 2 \over 2 x^3 (2x+3)^2}.
 \end{split}
\end{equation}
Since $x>0$ and $a>0$, the stability condition for the vacuum is automatically satisfied for this model (\ref{K1M}).

\subsection{The Probability Distribution of $\Lambda$}
The probability distribution $P(\Lambda)$ is easy to obtain :
\begin{equation}
 P(\Lambda) = \int_{-1}^0 dy \, P(y) \, \delta \left(\frac{a^3A^2e^{3+2 X}}{9+6 X} - \Lambda  \right)
 \label{P(Lambda0)}
\end{equation}
where $P(y)$ is the distribution of $-1< y \leq 0$ and $X(y) = {\cal W}_{-1} (3 e^{-3/2} y/2)$.
If we set $A=1$ for simplicity, $y$ obeys the same distribution as that of $c_1$.
The resulting probability distribution $P(\Lambda)$ is easy to evaluate numerically.

It is interesting to see analytically the peaking behavior of $ P(\Lambda)$. Given $P(y)$, the integration in (\ref{P(Lambda0)}) can be easily performed using the property of delta function $\delta ( g(y) ) = {|g'(y_0)|^{-1}} \delta (y-y_0)$. So we need to express $y$ as a function of $\Lambda$. That is, let us solve (\ref{Lambda}) for $y$.
With $a = A = 1$, let us choose a uniform distribution for $-1 < y \le 0$.
Introducing $p = -(3+2 X)$, (\ref{Lambda}) becomes
\begin{equation}
  - 1/3 \Lambda = p \, e^{p}.
\label{pequation}
\end{equation}
Now the physical constraint for $X(y)={\cal W}_{-1} (z)$ as in (\ref{solution for T}) requires $p$ to satisfy
$ X  = - (p+3)/2 < -{3/2}$.
So the solution for $p>0$ is given in terms of the other branch of the Lambert ${\cal W}$ function, namely,
\begin{equation}
  p = {\cal W}_0 \left( - {1 \over 3 \Lambda}\right)
  \stackrel{\Lambda \sim 0^-}{\rightarrow} -\ln\left[3 \Lambda \ln \left(-3 \Lambda\right)\right] + \cdots
  \label{P(Lambda)}
\end{equation}
Rewriting (\ref{Lambda}) using the basic relation of the Lambert ${\cal W}$ function, we have
\begin{equation}
 \Lambda = \left({3\over 2} e^{-3/2} y\right)^2 {e^3 \over X^2 (9 + 6 X)},
\end{equation}
Since now $X(y)$ is related to $p= {\cal W}_{0} (-1/3\Lambda)$, we obtain, after using (\ref{pequation}) and rewriting (\ref{Lambda}) using the basic relation of the Lambert ${\cal W}$ function,
\begin{equation}
 y = -{1\over 3} e^{-{\cal W}_0(-1/3\Lambda)/2} \left[{\cal W}_0 \left(- {1\over 3 \Lambda}\right) + 3\right].
  \label{y solution}
\end{equation}

Using (\ref{y solution}), we get, after some calculations,
\begin{align}
 P(\Lambda) =& {1\over 2} \left[3+2 {\cal W}_{-1} \left(f(\Lambda)\right) \right]^2 e^{-3/2 - {\cal W}_{-1}\left(f(\Lambda)\right)}, \label{result of P(Lambda)} \\
 f(\Lambda) =& - {1\over 2} \left[3 + {\cal W}_0 \left(-{1\over 3\Lambda}\right) \right] e^{-3/2 - {\cal W}_0 (-1/3\Lambda)/2}.\notag
\end{align}
Using the expansion formulae of the ${\cal W}$ functions, we find that $P(\Lambda)$ is actually divergent as $\Lambda \rightarrow 0^-$,
\begin{equation}
 P(\Lambda) \stackrel{\Lambda \rightarrow 0^-}{\simeq} {(3+\ln |\Lambda|)^2 \over 2 e^{3/2} \sqrt{|\Lambda|}} + \cdots.  
\end{equation}
We present both the ``analytical result'' (\ref{result of P(Lambda)}) and the numerical result (\ref{P(Lambda0)}) in figure\ref{fig:comparison-lambda}, where $a=A=1$ and $-1 \le c_1 \le 0$ has an uniform distribution.
Note that $P(\Lambda)$ is properly normalized in this paper, i.e., $\int P(\Lambda)d\Lambda=1$.

\begin{figure}
 \begin{center}
  \includegraphics[width=24em]{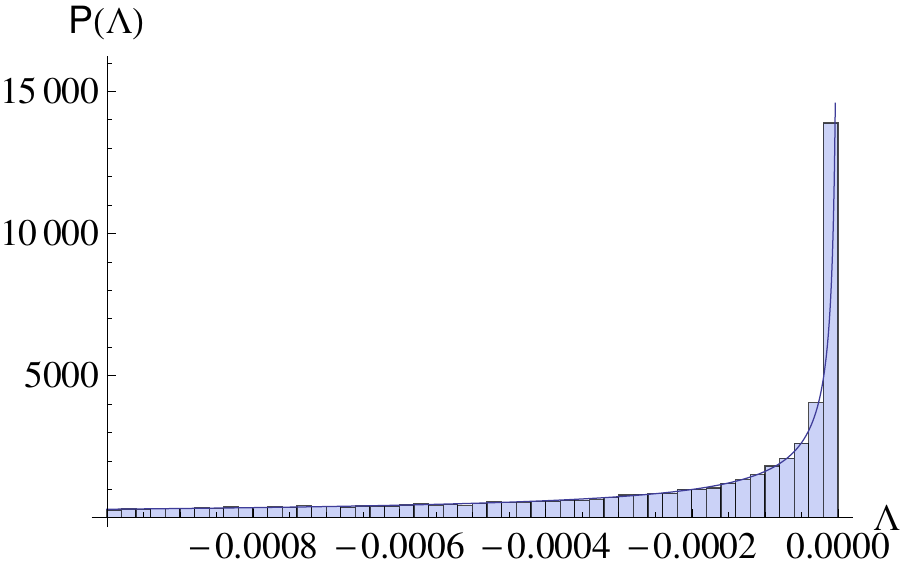}
 \end{center}
 \caption{The probability distribution $P(\Lambda)$ of the single K\"ahler modulus model (\ref{K1M}). The (blue) bars are numerical result for  (\ref{P(Lambda0)}) while the curve is for the analytical formula (\ref{result of P(Lambda)}).
 Both agree nicely even at $\Lambda = 0^-$, where $P(\Lambda)$ is divergent.}
 \label{fig:comparison-lambda}
\end{figure}

\section{Solving the Model}

\subsection{Solutions}
Now we are ready to solve the model (\ref{total potential}) with $h^{2,1}$ number of complex structure moduli for supersymmetric vacua, i.e., $D_T W=D_S W = D_{U_i} W = 0$. In order to find the properties with a large number of $U_i$, some semi-anayltic solution is crucial. So we restrict ourselves to real flux parameters $b_i$ $c_i$ and $d_i$ (thus cutting the number of random parameters by half). Picking the $\im T=\im S = \im U_i=0$ solution, we have 
\begin{subequations}
 \begin{align}
  &u_i = {-1\over h^{2,1} -2} {{\hat c}_1 - s c_2 \over b_i - s d_i}, \\
   &(h^{2,1} -2) {{\hat c}_1 + s c_2 \over {\hat c}_1 - s c_2} =  \sum_{i=1}^{h^{2,1}} {b_i + s d_i \over b_i - s d_i},\\
  & 3W+2xAe^{-x}=0, \\
  &W = - {2 ({\hat c}_1 + s c_2) \prod_{i=1}^{h^{2,1}} (b_i - s d_i) \over  \sum_{i=1}^{h^{2,1}} (b_i + s d_i) \prod_{j\neq i} (b_j - s d_j)}
 \end{align}
 \label{solution of complex sector}
\end{subequations}
where $s = \re S, u_i = \re U_i$, $x=a \re T=at$ and ${\hat c}_1= c_1 + Ae^{-x}$.
Note that, with (\ref{solution of complex sector}d), (\ref{solution of complex sector}b,c)
form a pair of equations for $s$ and $x$. Solving them determines $s$ and $x$ (as well as $u_i$), so the minimum $W$ (\ref{solution of complex sector}d) becomes a function of the parameters of the model. 

To solve this efficiently, we assume that the complex structure and dilation sector are stabilized at higher energy scales than that of the K\"ahler modulus $t$.
In this hierarchical setup (to be checked a posteriori), we first stabilize $u_i$ and $s$ in the model (\ref{total potential}) and then the K\"ahler modulus stabilization can be dealt with. Under ${\hat c}_1= c_1 + Ae^{-x} \rightarrow c_1$, $W \rightarrow W_0$, and (\ref{solution of complex sector}a,b,d)
reduce to those studied
in \cite{Sumitomo:2012vx} based on \cite{Rummel:2011cd}.
Although we deal with approximate solutions here, we can solve for all $u_i$ even in the presence of the non-perturbative term $Ae^{-x}$. After determining $s$ (which has multiple solutions) and $W_0$, we solve for $x$ and the minimum $W$.

\subsection{Probability Distribution of $\Lambda$}
To solve for the remaining K\"ahler modulus $x=at$, we simply replace $c_1$ in the earlier analysis by the solved $W_0$. 
The $\Lambda$ of the supersymmetric vacuum is now determined to be 
\begin{equation}
 \Lambda = -3 e^K |W|^2 = {1\over 2^{h^{2,1}+1} s \prod u_i } { a^3 A^2 e^{3 + {\cal W}_{-1} (z)} \over 9 + 6 {\cal W}_{-1} (z)}.
\end{equation}
It is now straightforward to  calculate probability distribution $P(\Lambda)$ numerically, given the distributions of the parameters of the model (\ref{total potential}). We see that $P(\Lambda)$ typically peaks sharply at $\Lambda =0^-$.

To be specific, consider the following scenario.
To deal with a divergent value of $\Lambda$ coming from $1/u_i$, we introduce a cutoff $f$ for flux inputs, $b_i=-f, -f \leq d_i \leq f$ while keeping $-1 \leq c_j \leq 1$ at each $h^{2,1}$ such that $90\%$ of the $\Lambda$ values
fall within the Planck scale ($M_P = 1$) range, similarly to \cite{Sumitomo:2012vx}.
Note that the qualitative result will not change even if  $b_i$ is randomized, though this will take more computer time.
The remaining $10\%$ of the data sitting over the Plank scale are discarded to maintain the validity of this supergravity approximation.
As before, we set $a=A=1$. 
The result is shown in figure \ref{fig:k+cs-distribution}. We see that the likely $|\Lambda|$ drops exponentially as $h^{2,1}$ increases, even though $\left< |\Lambda| \right>$ stays more or less constant. Similar qualitative properties are expected for some choices of distributions for the parameters, but not for others.

What happens when we turn on a supersymmetry breaking term? If this term is simply an additive term to the potential $V$ in (\ref{total potential}) and results in an additive term to $\Lambda$, then the peaking behavior of $P(\Lambda)$ will generically be erased \cite{Sumitomo:2012wa}. To maintain and even enhance the peaking behavior in $P(\Lambda)$, the supersymmetry breaking term must couple to the rest of $V$. So a good understanding of the supersymmetry breaking dynamics of the KKLT scenario (including back-reactions) is crucial.

\subsection{Probability of positive mass matrix}
Since we consider basically supersymmetric moduli stabilization for both the complex structure and K\"ahler sectors, all eigenvalues of the mass-squared matrix are expected to satisfy the Breitenlohner-Freedman bound \cite{Breitenlohner:1982jf}. However, for the purpose of uplifting to de-Sitter vacua (where the uplifting will be exponentially small) later, it is motivating to investigate the probability that all of the eigenvalues are positive at the supersymmetric vacua. Let us consider the Hessian 
\begin{equation}
 H_{IJ} = \partial_{\psi_I} \partial_{\psi_J} V|_{\rm extremal},
 \label{Hess}
\end{equation}
where $\psi_I$ runs through $x, s, u_i$.
Note that the extremum solution is inserted after taking the derivatives.
Inserting the numerical solutions into the Hessian (\ref{Hess}), 
we can check the probability of positively defined mass squared matrix since the positivity of Hessian is the necessary and sufficient condition for the positivity of mass squared matrix at extremal points, due to the linear transformation.

\begin{table}
 \begin{center}
  \begin{tabular}{|c||c|c|c|c|c|c|}
   \hline
   $h^{2,1}$ & 1 & 5 & 10 & 15 & 20 & 25 \\
   \hline
   Probability & 0.897 & 0.981 & 0.984 & 0.989 & 0.990 & 0.994\\ \hline
  \end{tabular}
 \end{center}
 \caption{The probability of having a positive Hessian ($\partial_i \partial_j V$) at $h^{2,1} = 1, 5,10, 15, 20, 25$. 
 The probability is approaching unity as $h^{2,1}$ increases.
 }
  \label{fig:probability}
\end{table}

Now we calculate the probability of having a positive Hessian in our model assuming the hierarchical setup.
The result is shown in table \ref{fig:probability}.
This probability actually increases as $h^{2,1}$ increases, and the value at $h^{2,1} = 5$ is already very close to unity.
Note that this probability is insensitive to the cutoffs for $b_i, d_i$ since the cutoffs we took only change the scale of quantities, and does not touch the sign.

The reason is actually quite simple.
Before introducing the non-perturbative term in $W$ in (\ref{total potential}), we have $|D_T W|^2-3|W|^2=K^{T\bar{T}}D_{T}W_0D_{\bar{T}}W_0-3|W_0|^2=0$ and the no-scale structure yields
$V = e^K (|D_S W|^2 + |D_{U_i} W|^2)$ which is positively defined and bounded below.
Then the supersymmetric solutions sit at Minkowski vacua, and the eigenvalues of Hessian are by definition semi-positive since the potential is convex downward at supersymmetric points.
Next let us introduce the non-perturbative term.
The correction to the supersymmetric solutions for the complex structure sector is of order of
\begin{equation}
 0 \geq {A e^{-x} \over W_0}  = {1\over -(2x/3 + 1)} \sim { 3 \over 2 \  \ln \left( {- W_0 / A } \right) } > -1,
\end{equation}
where we used (\ref{susy condition}) and (\ref{solution for T}) for small $-W_0/A$, and $x >0$.
Note that the K\"ahler modulus stabilization (\ref{susy condition}) requires $0\leq -W_0/A < 1$.
In fact, the distribution of $W_0$ is more peaked toward $W_0 = 0$ as $h^{2,1}$ increases, therefore smaller values of $W_0$ become more likely \cite{Sumitomo:2012vx}.
Thus we see that the correction due to the non-perturbative term for K\"ahler modulus stabilization gets smaller and becomes negligible when we increase $h^{2,1}$. 
This is the reason why the positivity of the Hessian for large $h^{2,1}$ is satisfied at most of extremal points.
This also means that the hierarchical structure between the complex structure sector and the K\"ahler sector we have employed for effective analysis is actually reliable, as anticipated in \cite{Kachru:2003aw}.

\section{Discussions}

This behavior of the probability is compatible with that studied in \cite{Bachlechner:2012at}, since their Gaussianly suppressed probability result will be applicable to our model only if we increase the number of K\"ahler moduli (lighter fields)
to a large number. That a positive mass matrix is almost automatic for large $h^{2,1}$ is very encouraging in the search of meta-stable de-Sitter vacua in the KKLT models.

Combining with the result in \cite{Sumitomo:2012vx}, we have a clear statistical preference for a vanishingly small $\Lambda$, either positive or negative, emerges. This is a consequence of the non-trivial functional dependences of $\Lambda$ on the flux parameters, when probability theory is applied to these string theory scenarios. Treating all parameters as physical, the $\Lambda$ studied here should be the physical $\Lambda$ as well. 

Notice that we have a number of reasonable ways to implement the distributions for the flux parameters. Also we have to introduce cut-offs to ensure the validity of the supergravity approximation. One may wonder whether these inputs have unwittingly forced us to have only very small $\Lambda$. This is an important question requiring further study. So it is somewhat reassuring in this preliminary study that the average $\Lambda$ stays within a few orders of magnitude of the Planck scale, suggesting that relatively large $\Lambda <1$ has not been thrown out by the various cut-offs introduced, which are necessary for the validity of the model.

A better understanding of the functions $C_i(U_j)$ (\ref{Ccoef}) as well as the distributions of the flux parameters will lead to a better determination of the likely values of $\Lambda$. It will be very interesting to find out what the statistical preference of $\Lambda$ will be when additional interactions are incorporated into the model (\ref{total potential}), or when other stringy scenarios are considered. The present work opens the door for further fruitful studies.

To conclude, let us make a few general optimistic remarks :
\begin{itemize}
 \item Our analysis shows that most meta-stable vacuum solutions are crowding around $\Lambda=0$ (at least in the corner of the landscape described by Type IIB). When positive non-supersymmetric vacua are included, the peaking at  $\Lambda=0$ behavior of $P(\Lambda)$ happens on both the positive and the negative sides of $\Lambda$. This property has to do with statistics rather than with energetics, since the latter would prefer more negative values of $\Lambda$. That our universe sits on the positive side may be historical. During the inflationary epoch, the universe has a relatively large positive vacuum energy density that drives inflation. As it rolls down the potential towards $\Lambda=0$, it may get trapped at the positive side of $\Lambda=0$ before it has a chance to reach the negative side of $\Lambda=0$. 
 \item Since the number of positive $\Lambda$ vacua away from $\Lambda=0$ are relatively very few, the universe may not get trapped in a relatively large $\Lambda$ de-Sitter vacuum, so eternal inflation is avoided.
 \item As we see that field theory itself has so far failed to provide an explanation of the smallness of $\Lambda$ without fine-tuning while string theory seems able to provide at least a statistical explanation, we like to boldly suggest that the very small but non-zero value of $\Lambda$ may be treated an experimental evidence for string theory.
\end{itemize}

\section{Acknowledgements}

We benefitted from stimulating discussions with X. Chen, L. McAllister, F. Quevedo, Y. Wang, and T. Wrase.

\bibliographystyle{utphys}
\bibliography{C:/Users/lunatic/Dropbox/papers/myrefs}

\end{document}